\documentclass[10pt]{article}
\usepackage{mathrsfs}
\usepackage{amsmath,amsfonts,amssymb,mathtools}
\usepackage{cite}
\usepackage{dsfont}
\usepackage{graphicx}
\usepackage{hyperref}
\usepackage{verbatim}
\usepackage{slashed}
\usepackage[all]{xy}
\usepackage{xcolor}
\usepackage{subfig}
\usepackage{tikz}
\usetikzlibrary{shapes.geometric,arrows}
\usepackage{relsize}
\usepackage{caption}
\usepackage{float}
\usetikzlibrary{positioning}
\usepackage{geometry}
\usepackage[T1]{fontenc}
\usepackage{lmodern}
\usepackage{stackengine} %per \ocirc 
 
\usepackage[utf8]{inputenc}

\hypersetup{
  colorlinks,
  citecolor=violet,
  linkcolor=blue,
  urlcolor=blue}

\csname @addtoreset\endcsname{equation}{section}
\DeclareMathAlphabet\mathbfcal{OMS}{cmsy}{b}{n}

\newcommand{\beq}{\begin{equation}}
\newcommand{\eeq}{\end{equation}}
\newcommand{\bea}{\begin{eqnarray}}
\newcommand{\eea}{\end{eqnarray}}
\newcommand{\ba}{\begin{array}}
\newcommand{\ea}{\end{array}}
\newcommand{\bit}{\begin{itemize}}
\newcommand{\eit}{\end{itemize}}
\newcommand{\nn}{\nonumber}

\newcommand{\mezzo}{\frac{1}{2}}
\newcommand{\complesso}{{\ \hbox{{\rm I}\kern-.6em\hbox{\bf C}}}}
\newcommand{\reale}{{\hbox{{\rm I}\kern-.2em\hbox{\rm R}}}}
\newcommand{\uno}{ \,  \raisebox{+0.14em}{{\hbox{{\rm \scriptsize ]}} \raisebox{-0.2em}{\kern-.8em\hbox{1}}}} \, }  %  operatore identit\`a

\newcommand{\p}{\partial}
\renewcommand{\a}{\alpha}
\renewcommand{\b}{\beta}
\newcommand{\g}{\gamma}

\newcommand{\D}{\Delta}
\newcommand{\e}{\epsilon}

\renewcommand{\l}{\lambda}
\renewcommand{\L}{\Lambda}

\newcommand{\m}{\mu}

\newcommand{\n}{\nu}
\renewcommand{\r}{\rho}
\newcommand{\s}{\sigma}

\begin{document}

%\begin{comment}

\begin{titlepage}

\vspace{0.3cm}

\begin{flushright}
%$IFUM$--1105--$FT$ \\
$LIFT$--14-2.26
\end{flushright}

\vspace{0.7cm}

\begin{center}
\renewcommand{\thefootnote}{\fnsymbol{footnote}}
\vskip 1mm  
{\Huge \bf             Reissner-Nordstrom in the   \\
\vskip 5mm
                      Bertotti-Robinson-Bonnor-Melvin \\
\vskip 9mm
                           electromagnetic field
}
\vskip 28mm
%Black holes in rotating, electromagnetic backgrounds and topological Kerr-Newman-NUT spacetime
%Black holes in the external Bertotti-Robinson-Bonnor-Melvin electromagnetic field
%Rotating and swirling binary black hole system balanced by its gravitational spin-spin interaction
%Most general Type-D Black Hole and the Accelerating Reissner-Nordstrom-NUT-(A)dS solution
%Ultrarelativistic boost of a black hole in the magnetic universe of Levi-Civita--Bertotti--Robinson

{\large {Marco Astorino$^{a}$\footnote{marco.astorino@gmail.com}
% Matilde Torresan$^{b}$\footnote{matilde.torresan@studenti.unimi.it}
}}\\
\renewcommand{\thefootnote}{\arabic{footnote}}
\setcounter{footnote}{0}
\vskip 8mm
\vspace{0.2 cm}
{\small \textit{$^{a}$Laboratorio Italiano di Fisica Teorica (LIFT),  \\
Via Archimede 20, I-20129 Milano, Italy}\\
} \vspace{0.2 cm}
%{\small \textit{$^{b}$Istituto Nazionale di Fisica Nucleare (INFN), Sezione di Milano \\
%Via Celoria 16, I-20133 Milano, Italy}\\
%} 
%\vspace{0.2 cm}
%
%{\small \textit{$^{b}$Universit\`a degli Studi di Milano}} \\
%{\small {\it Via Celoria 16, I-20133 Milano, Italy}\\}

\end{center}

\vspace{4.9cm}

\begin{center}
{\bf Abstract}
\end{center}
{The magnetic Reissner-Nordstrom black hole embedded into the Bertotti-Robinson magnetic field is presented. Its type is computed to belong to the Petrov D. Then it is used, as a seed, to generate, through the Harrison transformation of the Ersnt's equations, the Reissner-Nordstrom metric in the Bertotti-Robinson-Bonnor-Melvin magnetic field. \\
Special cases generalising, the vacuum hairy type I Schwarzschild solution is discussed.\\
The local equivalence between various forms of the Schwarzschild-Bertotti-Robinson is shown by exhibiting the explicit diffeomorphism.}

\end{titlepage}

\addtocounter{page}{1}

\newpage

%\tableofcontents
%\newpage

\section{Introduction}
\label{sec:introduction}

Astrophysical black holes, in particular the ones at the center of galaxies, are usually surrounded by huge magnetic fields. In general relativity the most famous and simple magnetic fields are the one discovered by Levi-Civita \cite{levi-civita-BR}, called Bertotti-Robinson \cite{bertotti},\cite{robinson} and the one found by Bonnor \cite{bonnor} and subsequently studied also by Melvin \cite{melvin}. Recently it has been shown in \cite{bh+BRBM} that these two apparently different solutions actually represent the same electromagnetic field, although embedded in geometries with slightly different symmetries. Since the two electromagnetic fields are basically the same they can be superposed or, by a fine tuning procedure, even subtracted from the spacetime, transforming an electro-vacuum solution into a vacuum, Ricci flat, spacetime. For instance, it is possible to generate new non-trivial black holes, in the Einstein theory, such as Schwarzschild with a gravitational hair \cite{static-typeI-bh} by mutually eliding the Bertotti-Robinson and Bonnor-Melvin electromagnetic field from the Schwarzschild black hole embedded in the Bertotti-Robinson and Bonnor-Melvin external magnetic field. This mechanism is completely general, so it works seamlessly also in different contexts: in presence of rotation or acceleration as observed in \cite{bh+BRBM}, \cite{static-typeI-bh}. In fact some of these straightforward extensions have been explicitly built in \cite{ma-li} or \cite{acc-hairy}. There is a less trivial generalisation of these spacetimes which, indeed, has not been built yet. It involves black holes endowed with the electromagnetic monopolar charge. The difficulty comes from the fact that: $(i)$ the interaction between the black hole electromagnetic field with the external Maxwell field makes the seed solution quite involved, $(ii)$ a clean solution for the Reissner-Nordstrom in the simple coordinate system of \cite{kerr-bertotti} is missing, at the moment, and that $(iii)$ the demagnetising constraint that remove only the external magnetic field in the presence of intrinsic electromagnetic charge is more subtle. Nevertheless, even if this task is more challenging, some progress in this direction can be done. These are the main purpose of this article: (I) to superpose the Bonnor-Melvin electromagnetic field to the Reissner-Nordstrom black hole to generalise the Schwarzschild-Bertotti-Robinson-Bonnor-Melvin of \cite{bh+BRBM} and (II) to remove the external magnetic field from the Reissner-Nordstrom-Bertotti-Robinson-Bonnor-Melvin solution to build a natural extension of the metric describing a Schwarzschild embedded in an external gravitational field \cite{static-typeI-bh}, that is a static hairy charged black hole.\\
The starting point is the choice of the seed and here the first difficulties emerge. In fact there is not a clear version of the Reissner-Nordstrom-Bertotti-Robinson spacetime in the coordinates of \cite{kerr-bertotti}. There is a hybrid metric containing both the acceleration and the monopolar charge in \cite{RN-BR0}, but it is not apparent how to switch off the acceleration in order to remain with a Reissner-Nordstrom black hole immersed into the Bertotti-Robinson spacetime\footnote{A recent intent where it is used as a seed is done in \cite{di-pinto}.}. Therefore the only seed metric available for our purpose is the Reissner-Nordstrom-Bertotti-Robinson solution recently found by Alekseev \cite{alekseev-RN-bertotti}. However for our procedure this solution is problematic for three reasons: (1) It is not clear, according to the actual literature, if it belongs to a branch of solutions equivalent with the one described by \cite{kerr-bertotti} or \cite{carminati}, not even in the simpler case: the Schwarzschild-Bertotti-Robinson black hole. Actually, while the Petrov class for both versions of the Schwarzschild-Bertotti-Robinson is the same ($D$), and both posses the limits to the Schwarzschild and Bertotti-Robinson solution, it is not known a diffeomorphism that can prove at least a local equivalence. (2) It is expressed in a parametrization and a coordinate system that is quite involved and it is not favourable for exploiting the symmetries transformations of the Ernst equations. (3) Its electromagnetic field is purely electric, while from a phenomenological point of view, the external magnetic field is physically more realistic and (4) from a theoretical point of view, for adding external fields, it is more convenient\footnote{The practical reason is that the Harrison and Ehlers transformations to add external electromagnetic or external gravitational fields to a black hole solution can become, in the simpler cases, just algebraic and no integration is needed.} to have the external Maxwell field of magnetic type. \\
In this paper we aim to address and clarify all these points before passing to the construction of the new solution through the Harrison transformation, in section \ref{sec:embedding-RN-BR-in-BM}. So we will focus in writing the magnetic Reissner-Nordstrom-Bertotti-Robison in a good parametrization and coordinate system, in sections \ref{sec:RN-BR-weyl}, \ref{sec:RN-BR-spherical}, and to test its coherence with the other black holes in the Bertotti-Robinson electromagnetic field in \ref{sec:equivalence}. There we will study some fundamental features of the new generated solution and we remove the external electromagnetic field in \ref{removal}.\\

\section{Reissner-Nordstrom black hole immersed in the\\ Bertotti-Robinson magnetic field}
\label{sec:RN-BR}

\paragraph{Setting} In this article we are dealing with the Einstein-Maxwell theory, whose field equations for the metric $g_{\m\n}$ and vector potential $A_\m$ are given by
\bea \label{Einsten-eq}
           & & R_{\m\n} -\mezzo R g_{\m\n} + \L g_{\m\n} = 2 \left( F_{\m\r} F_{\n}^{\ \r} - \frac{1}{4} F_{\r\s} F^{\r\s} \right)  \ ,\\
           & & \p_\m \left(\sqrt{-g} F^{\m\n} \right) = 0 \ ,\label{Maxwell-eq}
\eea
where the electromagnetic field $F_{\m\n}$ is defined as usual: $F_{\m\n} = \p_\m A_\n - \p_\n A_\m$ . All spacetime considered here are axisymmetric and stationary, therefore they posses at least a couple of commuting Killing vectors $\partial_t, \partial_\varphi$ so functions and fields depend only on a couple of non-Killing coordinates\footnote{The remaining two non-Killing coordinates we will use here are the Weyl or the spherical ones $(\alpha, \beta)$, ($r,x$).}.    \\

\subsection{RN-BR in Weyl coordinates ($\alpha, \beta$)} 
\label{sec:RN-BR-weyl}

As a starting point we will use a variation of the Reissner-Nordstrom solution embedded in the Bertotti-Robinson electric field found in \cite{alekseev-RN-bertotti}, that can be written in Weyl coordinates ($\a,\b$) as
\beq \label{metric-alpha-beta}
        ds^2 = g_{tt}(\a,\b) dt^2 + \frac{g_{\a\a}(\a,\b) \ \hat{b}^2 \ \left( d\a^2 + d\b^2  \right) }{\sqrt{\left[\hat{b}^2+\a^2+(\b-\b_1)^2 \right]^2 - 4 \hat{b}^2 \a^2}}  - \frac{\a^2}{g_{tt}(\a,\b)} d\varphi^2 \ , 
\eeq
with
\bea
         g_{tt} &=& - \frac{(\hat{b}^2+x_1^2)(x_2^2-\s^2)[x_2+\m y_1 + \n(x_1+\m y_2)]^2}{\hat{b}^2[x_2^2-\s^2-\n^2(\hat{b}^2+x_1^2)]^2}  \ , \nn \\
  g_{\a\a} &=&  \frac{f_0 D^2}{x_2-\s^2 y_2} \ ,  \nn \\
  x_1 &=& \frac{\b - \b_1}{y_1} \ , \hspace{1.4cm}  y_1 \, = \,  - \frac{\sqrt{\hat{b}^2-\a^2 - (\b - \b_1)^2 +\sqrt{[\hat{b}^2+\a^2+(\b - \b_1)^2]^2-4\hat{b}^2\a^2}}}{\hat{b} \sqrt{2}}  \ , \nn \\
  x_2 &=& \frac{1}{2} \left( R_+ + R_-\right)  \ ,  \hspace{0.4cm}   y_2 \, = \, \frac{1}{2\s} \left( R_+ - R_-\right) \ , \nn \\
  R_\pm &=& \sqrt{\a^2+(\b  -\b_1 - \ell \pm \s)^2}   \ , \nn \\
  \s &=& \sqrt{\m^2+2\ell\m\n-(\hat{b}^2+\m^2) \n^2} , \hspace{1cm} \ell \ = \ \b_2 - \b_1 \ ,\label{elle}
\eea
\beq D = [\ell^2+(\hat{b}^2+\m^2)(1-\n^2)] [x_2+\m y_1 + \n (x_1+\m y_2)] -2[\ell \n + \m(1-\n^2)][\ell(x_1-\m y_2) + (\hat{b}^2 + \m^2)(y_1+\n y_2)] , \nn
\eeq
%\bea
%         D &=& [\ell^2+(\hat{b}^2+\m^2)(1-\n^2)] [x_2+\m y_1 + \n (x_1+\m y_2)] -2[\ell \n + \m(1-\n^2)][\ell(x_1-\m y_2) + (\hat{b}^2 + \m^2)(y_1+\n y_2)] ,\nn \\
          %g_{\a\a} &=& [\n(\hat{b}^2+\m^2)-2\ell \m - \ell^2\n]x_2 + (\hat{b}^2+\ell^2-2\ell\m\n+\s^2)x_2 + [\hat{b}^2(\m+2\ell\n)+\m(\s^2-\ell^2)] y_1 + [] y_2 \\
         % g_{\a\a} &=&   \left(b^{2}+\ell^{2}-2\ell\mu\nu+\sigma^{2}\right) \left(x_{2}+\nu x_{1}+\mu y_{1}+\mu\nu y_{2}\right) -2\left(\mu+\ell\nu-\mu\nu^{2}\right) \left[ \ell(x_{1}-\mu y_{2}) + (b^{2}+\mu^{2})(y_{1}+\nu y_{2}) \right] , \nn
%\eea
where the only component of the vector electromagnetic potential is 
\beq \label{Atalphabeta}
         A_t = - \frac{\n(\hat{b}^2+x_1^2)(x_2+\m y_1)+(x_2^2-\s^2)(x_1+\m y_2)}{\hat{b}(x_2^2-\s^2-\n^2(\hat{b}^2+x_1^2))} \ .
\eeq

The physical parameters $\m,\, \n,\, \hat{b}$ are related to the mass, the intrinsic electric charge of the black hole, the intensity of the external Bertotti-Robinson electromagnetic field, while $\ell,\, \b_1,\, \b_2$ are geometrical properties related to the position of the black hole on the axis of symmetry and the relative distance between the black hole and the Bertotti-Robinson field. A relation between the physical parameters and the physical charges such as the mass $m$ and the electric charge $e$ has been also proposed in \cite{alekseev-RN-bertotti}
\beq
          \m = m - \frac{e \ell}{-e+\sqrt{\hat{b}^2+(m+\ell)^2}} \ , \hspace{2cm} \n = \frac{e}{\sqrt{\hat{b}^2+(m+\ell)^2}} \ .
\eeq
In terms of the physical parameters $\s=\sqrt{m^2-e^2}$.\\

The analysis of possible conical singularities of the spacetime (\ref{metric-alpha-beta})-(\ref{elle}) can be done computing the possible deficit or excess angle defined as
\beq
      \lim_{\a \to 0} \  \frac{g_{\varphi\varphi}}{\a^2 g_{\a\a}}
\eeq
on the two disconnected regions outside the event horizon: from the north and the south pole, for $\b>\b_2+\s$ and $\b<\b_2-\s$ respectively. Imposing the unitary value of the above limit in the two different sectors, we get a couple of regularity conditions for the metric parameters
\beq
      \frac{\hat{b}^2}{f_0\{\hat{b}^3(1+\n^2) \mp 2\sqrt{\hat{b}^2} [\hat{b}^2\n + \m(\b_1 -\b_2 +\m \n)] + \hat{b} [\m^2+(\b_1-\b_2+\m\n)^2] \}} = 1 \ ,
\eeq
which, in terms of the physical parameters, become
\bea
      1 &=&    \frac{1}{f_0 \left[e - \sqrt{\hat{b}^2+(m+\ell)^2} \right]^4}      \ , \\ \nn
      1 &=&      \frac{\left[e- \sqrt{\hat{b}^2+(m+\ell)^2} \right]^4}{f_0 \left[\hat{b}^4+(\ell^2-\s^2)^2+2\hat{b}^2(\ell^2+\s^2) \right]^2}      \ .
\eea
The equilibrium configuration, for non-trivial finite physical parameters and without the necessity of postulating the presence of singular matter in the form of a string or a strut, needs $\ell \neq 0$. The $f_0$ parameter is basically a gauge parameter of the metric and can be set to comply with the first constraint, such as
\beq
      f_0 = \left[e - \sqrt{\hat{b}^2+(m+\ell)^2} \right]^{-4} \ .
\eeq
Hence the second constraint become 
\beq  \label{conditio}
       \frac{\left[e - \sqrt{\hat{b}^2 + (m + \ell)^2}\right]^4}{\left[\hat{b}^2 + (\ell - \sigma)^2\right] \left[\hat{b}^2 + (\sigma + \ell)^2\right]} = 1 \ .
\eeq
This condition, with the setting of the standard range of the azimuthal angle $\varphi \in [0,2\pi) $, can be fulfilled for $\ell\neq 0$, by setting one of physical parameters to a proper finite real non-trivial value. If the condition is fulfilled by unbounded values of the parameters, the range of $\varphi$ might be different from the standard one.

The study of the algebraic class of the spacetime, following \cite{stephani-big-book}, \cite{Type-I}, points out that the charged black hole in the Bertotti-Robinson electromagnetic field belongs to the $D$ class.\\  

Since we want to superpose the magnetic Bonnor-Melvin electromagnetic field to the Bertotti-Robinson magnetic field the first step is to rotate the electric field of \cite{alekseev-RN-bertotti} into the magnetic field. This procedure can be done is several ways, through the unitary restriction of the $(I)$ transformation of the Ernst equations, as described in \cite{enhanced} or similarly thanks to the electromagnetic duality of the Maxwell field in four-dimensions, basically using the fact that the Hodge-dual $\tilde{F}_{\m\n} := \frac{1}{2} \e_{\m\n\r\s} F^{\r\s}$ of a given Faraday matrix $F_{\m\n}$ produces the same stress-energy tensor of the initial electromagnetic field (i.e. $T_{\m\n}[F]=T_{\m\n}[\tilde{F}]$), therefore one gets a new solution with the same metric and a Hodge rotated electromagnetic field. In this case the Hodge dual of the purely electric solution \cite{alekseev-RN-bertotti} becomes a purely magnetic electromagnetic field. In practice, for a metric of the form (\ref{metric-alpha-beta}), if we want to obtain the magnetic component of the vector potential, $A_\varphi$, we need to solve the following system of partial differential equations
\bea
               \frac{\p}{\p \a} A_\varphi(\a,\b)      & = &  \frac{\a}{g_{tt}(\a,\b)} \ \frac{\p}{\p \b} A_t(\a,\b)   \ ,\\
               \frac{\p}{\p \b}  A_\varphi(\a,\b)     & = &  \frac{-\a}{g_{tt}(\a,\b)} \ \frac{\p}{\p \a} A_t(\a,\b)  \ .
\eea
Up to an arbitrary gauge constant $A_{\varphi0}$, we get
\beq \label{Aphi}
   A_\varphi = - \hat{b} \,  \frac{\m\,(1-\n^2)+(x_2+\n x_1)\,(y_1-\n y_2)}
{(x_2+\n x_1)+\m\,(y_1+\n y_2)} \ .
\eeq
The metric (\ref{metric-alpha-beta})-(\ref{elle}) is now solution of the Einstein and Maxwell field equations (\ref{Einsten-eq})-(\ref{Maxwell-eq}) with $A_t=0$ and $A_\varphi$ as in (\ref{Aphi}).\\
Note that after the duality transformation the physical interpretation of the parameters $\n$ (or $e$) and $\hat{b}$ passes from the intensity of the monopolar electric charge of the black hole and external electric field respectively to the intensity of the intrinsic monopole magnetic charge and of the external magnetic field. \\
Of course a solution composed by the metric (\ref{metric-alpha-beta})-(\ref{elle}), the combination of the electric and magnetic fields, as expressed in (\ref{Atalphabeta}) and (\ref{Aphi}), can be obtained by considering the following dyonic vector potential
\beq
                A_\m = \Big( \cos \g \ A_t , \ 0 , \  0 ,\  \sin \g \ A_\varphi  \Big) \ .
\eeq
Note however that this construction does not represent the most general Reissner-Nordstrom black hole immersed in the Bertotti-Robinson magnetic field. A stationary extension is possible. It has an extra integration constant in the electromagnetic potential that makes the intensities of the external fields and intrinsic charges of the black hole independent. It takes into account the interaction between the monopole charges and the external electromagnetic field that makes Reissner-Nordstrom rotate.\\

\subsection{Equivalence with the other form of the Schwarzschild-Bertotti-Robinson}
\label{sec:equivalence}

For $e=0$ (or equivalently $\n=0$) the monopolar charge of the black hole disappears and  we remain precisely with Schwarzschild embedded into the Bertotti-Robinson external magnetic field \cite{alekseev-garcia}. \\
However other forms of the Schwarzschild solution  inside the Bertotti-Robinson background, for instance the compact one of \cite{carminati}, \cite{kerr-bertotti}. Both families share, for zero mass parameter, the same limits to the Bertotti-Robinson background and to the Schwarzschild metric. Furthermore  both families are of type D, according to the Petrov classification \cite{ortaggio}, therefore they are likely to be diffeomorphic. At the moment, there is some confusing and misleading literature on that matter. Thus  it could be useful to shed some light on this open  question. \\

Consider the metric (\ref{metric-alpha-beta})-(\ref{elle}) supported by the magnetic field deduced by (\ref{Aphi}), with $\n=0$ and $\ell=0$, that is a special case of the solution \cite{alekseev-garcia}, where both the black hole and external electromagnetic field has a specific position, let's say at the origin of the coordinates. This is a special case also from a physical perspective: it is the configuration that is void of conical singularities. So this is the best candidate to test the equivalence with the Schwarzschild-Bertotti-Robinson solution in the representation of \cite{kerr-bertotti}
\beq \label{schwarschild-br-spherical}
      ds^2 = \frac{-\left( 1 -\frac{2\bar{m}}{\bar{r}} - \bar{m}^2B^2 \right) \left( 1+ B^2\bar{r}^2\right) dt^2 + \frac{d\bar{r}^2}{\left(1 - \frac{2\bar{m}}{\bar{r}} - \bar{m}^2B^2 \right) \left( 1+ B^2\bar{r}^2\right)} +\frac{\bar{r}^2 d\bar{\theta}^2}{\D_{\bar{\theta}}} + \bar{r}^2 \sin^2{\bar{\theta}} \D_{\bar{\theta}} \D_\varphi^2 \, d\varphi^2}{1+B^2\bar{r}^2-B^2\bar{r} \cos^2 {\bar{\theta}} (\bar{r}-2\bar{m}-\bar{r}B^2\bar{m}^2)} \ ,  \nn
\eeq
where\footnote{The constant $\D_\varphi$ can be set to $1/(1+B^2m^2)$ to avoid conical singularities.}
\beq
         \D_{\bar{\theta}}({\bar{\theta}}) = 1+\bar{m}^2B^2\cos^2{\bar{\theta}} \ .
\eeq
It can be proven that the following coordinate transformation, borrowed (apart from a sign) from \cite{kerr+bubble}, 
\bea \label{alpha-transf}
        \a(\bar{r},{\bar{\theta}}) &=& \frac{\sqrt{1+B^2 \bar{r}^2} \sqrt{\bar{r}^2-2\bar{m}\bar{r}-B^2 \bar{m}^2 \bar{r}^2} \sin {\bar{\theta}} \sqrt{1+B^2 \bar{m}^2 \cos^2 {\bar{\theta}}}}{1+B^2[\bar{r}^2-(\bar{r}^2 - 2\bar{m}\bar{r}-B^2\bar{m}^2\bar{r}^2)\cos^2 {\bar{\theta}}]} \ , \\
          \b(\bar{r},{\bar{\theta}}) & = &   \frac{\cos{\bar{\theta}} (\bar{r}-\bar{m})(1+B^2\bar{m}\bar{r})}{1+B^2[\bar{r}^2 - (\bar{r}^2 - 2\bar{m}\bar{r} -B^2\bar{m}^2\bar{r}^2)\cos^2 {\bar{\theta}}]} \ , \label{beta-transf}
\eea
with the reparametrisation $\hat{b} = 1/B,\ \m=\bar{m}$, maps the Schwarzschild-Bertotti-Robinson of \cite{alekseev-garcia} (or  (\ref{metric-alpha-beta})-(\ref{elle}) with $\n=\b_1=\b_2=0$) into 
(\ref{schwarschild-br-spherical}).\\
Also the magnetic component of the gauge potential (\ref{Aphi}) transforms accordingly: up to a gauge constant it becomes 
\beq \label{sch-br-spherical-Aphi}
           A_\varphi = \frac{1+B^2\bar{m} \bar{r}\cos^2{\bar{\theta}}-\sqrt{1+B^2[\bar{r}^2 - (\bar{r}^2 - 2\bar{m}\bar{r} -B^2\bar{m}^2\bar{r}^2)\cos^2 {\bar{\theta}}]}}{B \sqrt{1+B^2[\bar{r}^2 - (\bar{r}^2 - 2\bar{m}\bar{r} -B^2\bar{m}^2\bar{r}^2)\cos^2 {\bar{\theta}}]}} \, \D_\varphi \ .
\eeq

So the two metrics describing Schwarzschild in the Bertotti-Robinson external electromagnetic field belong to the same family, although in different sets of coordinates.  \\
All that said, we recall that the range of the local equivalence, obviously, is constrained by the domain of the diffeomorphism. The two manifolds can be considered physically equivalent where the change of coordinates is well defined. More specifically looking at the determinant of the Jacobian of the transformation (\ref{alpha-transf})-(\ref{beta-transf}) 
\begin{equation}
\det J := 
\begin{vmatrix} \dfrac{\partial \alpha}{\partial \bar{r}} & \dfrac{\partial \alpha}{\partial \bar{\theta}} \\[12pt] \dfrac{\partial \beta}{\partial \bar{r}} & \dfrac{\partial \beta}{\partial \bar{\theta}} \end{vmatrix} = 
\end{equation}
$$ \frac{\left[ \bar{m}^2 (1 + B^2 \bar{m}\bar{r})^2 \cos^2\bar{\theta} - (\bar{r} - \bar{m})^2 \right] \left[ (1 + B^2 \bar{m} \bar{r})^2 + B^2 (\bar{r} - \bar{m})^2 \cos^2\bar{\theta} \right]}{\sqrt{1 + B^2 \bar{r}^2} \sqrt{\bar{r}^2 - 2\bar{m}\bar{r} - B^2 \bar{m}^2 \bar{r}^2} \sqrt{1 + B^2 \bar{m}^2 \cos^2\bar{\theta}} \; \left\{ 1 + B^2 \left[ \bar{r}^2 - (\bar{r}^2 - 2\bar{m}\bar{r} - B^2 \bar{m}^2 \bar{r}^2) \cos^2\bar{\theta} \right] \right\}^2}$$
we can infer that it is invertible outside the event horizon for $\bar{r}>2m/(1-B^2m^2)$ \footnote{Of course, when considering the full three-dimensional space, the coordinate transformation from the Weyl and the spherical coordinates are, as usual, not invertible also on the poles.}

\section{Reissner-Nordstrom in a Bertotti-Robinson-Bonnor-Melvin external magnetic field}
\label{sec:RN-BRBM}

We are interested in building a Reissner-Nordstrom solution including a superposition of both the Bertotti-Robinson and the Bonnor-Melvin electromagnetic field, a spacetime describing a black hole endowed with monopolar charge embedded inside a  general electromagnetic field. Hence we want to extend the black hole solution found in \cite{bh+BRBM} in the presence of the monopolar charge.\footnote{While the application of a charging transformation, such as the (electric) Harrison may add the monopolar charge to the system, it adds charges also asymptotically, so we  would add an extra electric charge also to the background, which, in that case, cannot be considered a proper Bertotti-Robinson-Bonnor-Melvin background, that is way this approach will not be discussed here. In any case the interested reader can find that solution in the appendix \ref{app:electric-harry}. The physical reason why it deforms also the background, if not trivial,  is that the external electromagnetic background can be thought as black holes very far away \cite{Emparan:2001gm}, at spatial infinity, the only trace of the solution remains in its electromagnetic field. Nevertheless the Harrison transformation acts and deforms on the monopolar and multipolar charges of the solution.} \\
To keep the model simple and to avoid stationary rotation and off-diagonal components, we need to consider electromagnetic fields of the same kind. In this case we will work with magnetic intrinsic charge for the RN black hole and magnetic external fields. Of course the same degree of complexity would have been achieved with the electric counterpart,  which can be obtained by the electromagnetic duality rotation of the final solution or by a (magnetic) Harrison transformation with an imaginary Lie parameter instead of real.\\

\subsection{Elettric or Magnetic RN-BR in spherical coordinates} 
\label{sec:RN-BR-spherical}

First of all, we  write the seed spacetime, the {\it Reissner-Nordstrom-Bertotti-Robinson} in spherical coordinates ($r, \,  x=\cos \theta $)
\beq \label{rn-br-rx-inizio}
        ds^2 = g_{tt}(r,x) dt^2 + g_{rr}(r,x) \left[ \frac{dr^2}{\D_r(r)} + \frac{dx^2}{\D_x(x)} \right] - \frac{\D_r(r) \D_x(x)}{g_{tt}(r,x)} d\varphi^2 \ , 
\eeq
where the ($r,x=\cos \theta$) are the spherical coordinates\footnote{Note that these do not coincide with the ($r,x$) coordinates of \cite{bh+BRBM}, which will be labelled as ($\bar{r},\bar{x}=\cos \bar{\theta}$) in the present paper. More about the differences between these two set of coordinates in appendix \ref{app:SBR-rx}.} and 
\bea
      g_{tt}(r,x) &=& -\frac{(1 + B^2 x_1^2) \left[ B e \sqrt{1 + B^2 m^2} (mx + x_1) + (1 + B^2 m^2) (-m + r + m y_1) \right]^2 \Delta_r}{\left[ e^2 + B^2 e^2 x_1^2 - (1 + B^2 m^2) \Delta_r \right]^2} \ , \nn \\
      g_{rr}(r,x) &=&  \frac{\left[ B e (mx - x_1) + \sqrt{1 + B^2 m^2} (m - r + m y_1) \right]^2}{\sqrt{4 B^2 (-1 + x^2) \Delta_r + \left[ 1 + B^2 \s^2 x^2 + B^2 \Delta_r \right]^2}}\ , \nn \\
      \D_r (r)  &=&  r^2 - 2 m r + e^2 \ , \nn\\
      \D_x(x) &=& 1-x^2 \ , \nn\\
      x_1 &=&  \frac{\sqrt{2} (m - r) x}{\sqrt{1 - B^2 \s^2 x^2 - B^2 \Delta_r + \sqrt{4 B^2 \left[ 1 + B^2 \s^2 \right] x^2 \Delta_r + \left[ 1 + B^2 \s^2 x^2 - B^2 \Delta_r \right]^2}}} \ ,\nn \\
      y_1 &=&  \frac{x_2 y_2}{x_1} \nn \\ 
      x_2  &=& r - m  \nn \\
      y_2 &=& x   \label{rn-br-rx-fine}   
\eea
The metric (\ref{rn-br-rx-inizio})-(\ref{rn-br-rx-fine}) can be supported by a magnetic field stemming, up to a constant gauge $A_{\varphi0}$ (that, in order to assure good $B \to 0$ limits can be chosen to include a $-1/B$ term), from $A_t=0$ and  
\beq \label{Aphirx}
A_\varphi (r,x) = 
 \frac{\left( B e x_1 - \sqrt{1 + B^2 m^2} x_2 \right) \left[ y_2 + B^2 x_1 (m + x_1 y_2 - m y_2^2) \right]}{B \sqrt{1 + B^2 m^2} x_1 \left\{1 + B^2 \left[ x_1^2 + m^2 (1 - y_2^2) \right] \right\}} \ ,
 \eeq
or alternatively by an electric field given by $A_\varphi=0$ and 
\beq \label{Atrx}
       A_t(r,x) = \frac{(mx + x_1) \left[ e \sqrt{1 + B^2 m^2} (1 + B^2 x_1^2) y_1 + B (1 + B^2 m^2) x \Delta_r \right]}{e^2 x (1 + B^2 x_1^2) - (1 + B^2 m^2) x \Delta r} \ .
\eeq
The dyonic combination of the magnetic and electric fields (\ref{Aphirx})-(\ref{Atrx}) can be considered also for the metric (\ref{rn-br-rx-inizio})-(\ref{rn-br-rx-fine}) by
\beq
                A_\m = \Big( \cos \g \ A_t , \ 0 , \  0 ,\  \sin \g \ A_\varphi  \Big) \ .
\eeq
$e$ and $B$ can be regarded as intensity of the intrinsic and external electric or magnetic field depending on the fact we are using the electric or magnetic field, respectively (\ref{Atrx}) or (\ref{Aphirx}).  \\
Notice that this solution is a subcase of the one of section \ref{sec:RN-BR}, where we have set $\b_1=\b_2=0$, so the relative position of the black hole with respect to the external field is characterised by $\ell=0$. The complete solution, in spherical coordinates, can be be obtained from the one of (\ref{metric-alpha-beta})-(\ref{Atalphabeta}) and (\ref{Aphi}), thanks to the standard change of coordinate between Weyl and spherical coordinates
\beq
         \a (r,x) = \sqrt{(r-m)^2-\s^2} \, \sqrt{1-x^2} \ \ , \hspace{1.3cm} \b (r,x) = \b_2+(r-m) \, x \ .
\eeq
In this frame of reference the limit of (\ref{rn-br-rx-inizio})-(\ref{rn-br-rx-fine}) to the Reissner-Nordstrom black hole are straightforward, just vanishing the external electromagnetic field $B \to 0$:
\beq \label{RN}
            ds^2 = - \left( 1- \frac{2 m}{r} + \frac{e^2}{r^2} \right) dt^2 + \frac{dr^2}{1- \frac{2 m}{r} + \frac{e^2}{r^2}} + r^2 d\theta^2 +r^2 \sin^2 \theta d\varphi^2 \ , 
\eeq
with $A_\m = (-e/r, 0, 0, 0) $ or $A_\m = (0 ,0,0, e \, x )$.\\
 Also the limit to the Bertotti-Robinson is clear, by removing the charged black hole setting $m=e=0$ 
\beq  \label{br-g-rx}
         ds^2 = - \frac{1}{2} \left( 1 + B^2 r^2 + H \right)  dt^2   + \frac{dr^2}{H} + \frac{r^2 dx^2}{(1-x^2) H} + \frac{1+B^2r^2-H}{2B^2} d\varphi^2 
\eeq
with  $ H(r,x) = \sqrt{(1-B^2 r^2)^2+4B^2r^2x^2}  $ and 
\beq \label{br-A-rx}
        A_\m = \left[ 0 , 0, 0 , \frac{H(r,x)}{\sqrt{2} B}  \right]
\eeq
It it easy to show that the above form of the Bertotti-Robinson can be mapped to the canonical conformally flat $AdS_2 \times S^2$ solution
\bea
        ds^2 &=& \frac{1}{B^2} \left( - \cosh^2 \chi d\tau^2 +d\chi^2 +d\vartheta^2 +\sin^2 \vartheta d\varphi^2 \right) \ , \nn \\
          A &=&  \frac{\cos \vartheta}{B} \, d\varphi \ ,
 \eea
by the coordinate transformation
\beq
      t = \frac{\tau}{B} \ \ , \hspace{1.1cm} r(\chi, \vartheta ) =  \frac{1}{B} \sqrt{\sinh^2 \chi + \sin^2 \vartheta}  \ \ , \hspace{1.1cm} x(\chi, \vartheta) = \frac{\sinh \chi \cos \vartheta}{ \sqrt{\sinh^2 \chi + \sin^2 \vartheta}} \  \nn .
\eeq
The Minkowski background is retrieved from (\ref{br-g-rx})-(\ref{br-A-rx}) when $B \to 0$; the limit is well defined.\\
On the other hand, if we set only $e=0$, we recover the ($\ell=0$) Schwarzschild Bertotti-Robinson \cite{alekseev-garcia}, but in spherical coordinates. \\
The electromagnetic monopole in the Bertotti-Robinson electromagnetic field is given by setting in (\ref{rn-br-rx-inizio})-(\ref{rn-br-rx-fine}) only $m=0$. \\

\subsection{Embedding RN-BR in the magnetic Bonnor-Melvin universe} 
\label{sec:embedding-RN-BR-in-BM}

Adding to the existing magnetic field of the Reissner-Nordstrom-Bertotti-Robinson solution the Bonnor-Melvin contribution is simple with the help of the magnetising Harrison transformation, as done in \cite{bh+BRBM} for $e=0$, the Schwarzschild-Bertotti-Robinson subcase. For details about the construction technique, in this notation,  see \cite{bh+BRBM}, \cite{enhanced}. The Harrison transformation is a Lie-point symmetry of the field equations therefore it introduces an extra continuous  parameter\footnote{The extra parameter in principle can be more general, i.e. it can be complex, corresponding to two real parameters. Usually the real part of the magnetising Harrison transformation adds external magnetic field, while the imaginary part adds external electric field. Here we consider for simplicity only the external magnetic field. Because of the Lorentz force a black hole endowed with magnetic intrinsic monopolar charge in the external electric field would rotate.}, $b$ characterising the intensity of the external Bonnor-Melvin magnetic field. The resulting metric can be written as 
\beq \label{rn-brbm}
        d\hat{s}^2 = \L^2(r,x)\left\{ g_{tt}(r,x) dt^2 + g_{rr}(r,x) \left[ \frac{dr^2}{\D_r(r)} + \frac{dx^2}{\D_x(x)} \right] \right\} - \frac{\D_r(r) \D_x(x) \D_\varphi}{g_{tt}(r,x) \L^2(r,x)} d\varphi^2 \ , 
\eeq
where
\bea
         \L(r,x) &=&  \frac{1}{4} \left[ ( b A_\varphi - 2)^2 -  \frac{b^2 \D_r \D_x}{g_{tt}} \right]  \ ,  \\
         \hat{A}_\m    &=&    \left\{ 0, 0 , 0, \frac{A_\varphi (2 - b A_\varphi) g_{tt} + b \D_r \D_x \D_\varphi}{2 g_{tt} \ \L}\right\} \ . \label{AphiBRBM}
\eea
We recall that, in (\ref{AphiBRBM}), $A_\varphi$ represents the seed magnetic potential (\ref{Aphirx}), afferent to Reissner-Nordstrom-Bertotti-Robinson, as above in this section. $\D_\varphi$ is introduced thanks to the scaling freedom of the Killing coordinate $\varphi$ to deal with possible conical singularity (notice that for sake of generality, also in the seed (\ref{rn-br-rx-inizio})-(\ref{rn-br-rx-fine}) it can be considered).

\paragraph{Limits} The interpretation of the solution (\ref{rn-brbm})-(\ref{AphiBRBM}) as the superposition of the Bertotti-Robinson-Bonnor-Melvin with the Reissner-Nordstrom black hole is quite straightforward. To convince oneself it is sufficient to look at some known limits:
\begin{itemize}
\item $b = 0$ :  Reissner-Nordstrom in Bertotti-Robinson magnetic field, the seed (\ref{rn-br-rx-inizio})-(\ref{Aphirx}).
\item $B = 0$ :  Reissner-Nordstrom in Bonnor-Melvin magnetic field magnetic field
%\end{itemize}
\beq \label{RN+BM}
        ds^2 =  \L_0^2(r,x) \left[ \left(1-\frac{2m}{r} + \frac{e^2}{r^2} \right) dt^2 + \frac{dr^2}{1-\frac{2m}{r} + \frac{e^2}{r^2}} + \frac{r^2 dx^2}{1-x^2} \right] + \frac{r^2(1-x^2) d\varphi^2}{\L_0^2(r,x)}
\eeq
\bea
\L_0(r,x) &=& 1- b\, e\, x + \frac{b^2}{4} \bigg[ r^2(1-x^2)  + e^2 x^2  \bigg] \ , \nn \\ 
A_\varphi &=&  \frac{4 e x - 2 b e^2 x^2 + 2 b r^2 (-1 + x^2)}{4 - 4 b e x + b^2 \left[ e^2 x^2 - r^2 (-1 + x^2) \right]} \ . \label{A-RN-BM}
\eea
This is one of the reasons why the solution in (\ref{rn-brbm})-(\ref{AphiBRBM}) can not belong to type-D, according to the Petrov classification. In fact even a subcase, the Reissner-Nordstrom in the Bonnor-Melvin universe, as in (\ref{RN+BM})-(\ref{A-RN-BM}) is known for being of the general type I.   

\item $ b = B = 0 $ : Reissner-Nordstrom (\ref{RN})

\item $ e = 0 $ : Schwarzschild in the Bertotti-Robinson-Bonnor-Melvin magnetic background, built in \cite{bh+BRBM} but in a different set of coordinates, see appendix \ref{app:SBR-rx}.

\item $ m = 0 $ : Magnetic monopole in the Bertotti-Robinson-Bonnor-Melvin magnetic background  

\item $m = e = 0 $ : Bertotti-Robinson-Bonnor-Melvin magnetic Background
\end{itemize}

We write explicitly the background in these coordinates 
\beq \label{BMMB-rx}
                       ds^2 =  f(r,x) \left[ - dt^2 + \frac{(2-S)}{(1-x^2)2B^2(1-S-B^2r^2)} \left( \frac{dr^2}{r^2} + \frac{dx^2}{1-x^2} \right) \right]  + \frac{r^2 (1-x^2)}{f(r,x)} d \varphi^2\
\eeq
with 
\bea
       f(r,x) &=& \frac{\left[4 B^2 - b^2 (\sqrt{2} \sqrt{S} - 2) - 2 b B (\sqrt{2} \sqrt{S} - 2)\right]^2 \left(S + 2 B^2 r^2 x^2\right)}{16 B^4 S}  \, \nn \\
       A_\varphi(r,x) &=& \frac{2 (b + B) (-2 + \sqrt{2} \sqrt{S})}{4 B^2 - b^2 (-2 + \sqrt{2} \sqrt{S}) - 2 b B (-2 + \sqrt{2} \sqrt{S})}  \ , \nn \\
       S(r,x) &=&  1-B^2r^2 + \sqrt{(1-B^2r^2)^2+4B^2r^2x^2} \label{BMMB-rx-fine}
\eea
From the study of the Kretschmann scalar invariant $\mathcal{K}=R_{\m\n\s\l} R^{\m\n\s\l}$  we deduce the above metric is regular, that is void from curvature singularities: $\mathcal{K}$ is everywhere bounded and continuous. Moreover the analysis of the axis of symmetry says the metric has no conical singularities. This is true also for the vacuum background recently discovered in \cite{bh+BRBM}, \cite{static-typeI-bh}, a subcase of the Bertotti-Robinson-Bonnor-Melvin background for $b=-B$. Using the more economical coordinates ($\bar{r},\bar{x}$) as in (\ref{schwarschild-br-spherical})-(\ref{sch-br-spherical-Aphi}), related to the ($r,x$) ones by
\beq \label{coorfinates-hat}
             \bar{r} = \sqrt{\mp \frac{-1+B^2r^2+\sqrt{1+B^4r^4-2B^2(1-x^2)}}{2B^2x^2}}  \ , \hspace{0.7cm}  \bar{x} = \frac{x}{r}\sqrt{\mp \frac{-1+B^2r^2+\sqrt{1+B^4r^4-2B^2(1-x^2)}}{2B^2x^2}}    \ ,  \nn
\eeq
 it reads 
\beq \label{ds-vacuum}
                       ds^2 = \left[\frac{1+\sqrt{1+B^2\bar{r}^2(1-\bar{x}^2)}}{2+ 2 B^2\bar{r}^2 (1-\bar{x}^2)} \right]^2 \left[-(1+B^2\bar{r}^2) \, dt^2 + \frac{d\bar{r}^2}{1+B^2\bar{r}^2} + \frac{\bar{r}^2 d\bar{x}^2}{1-\bar{x}^2} \right] + \frac{4\bar{r}^2 (1-\bar{x}^2) \ d\varphi^2}{\left[1+\sqrt{1+B^2\bar{r}^2(1-\bar{x}^2) }\right]^2}  \ .  
\eeq
In this coordinates the Kretschmann scalar is simply
\beq \label{k-vacuum}
                      \mathcal{K} = \frac{768 B^4 \left(1 + B^2 \bar{r}^2 (1 - \bar{x}^2)\right)^3}{\left[1 + \sqrt{1 + B^2 \bar{r}^2 (1 - \bar{x}^2)}\right]^6} \nn  \ .
\eeq
So it is easy to confirm its finiteness in these coordinates $(r,x)$ or $(\bar{r},\bar{x})$, however note that not all these coordinates cover the whole spacetime completely, therefore a proper analytical extension might be needed if we  want to consider this metric as a full and proper spacetime rather than a local model\footnote{Since the curvature scalar invariant are not divergent in $\bar{r}\to \infty$ it is possible to find an analytical extension for the $(t,\bar{r},\bar{x},\varphi)$ chart.}. In \cite{herdeiro} it is suggested that using another sets of coordinates the solution is supported by an equatorial annulus of delta-like matter, however being both sets of coordinates used in \cite{herdeiro} non-invertible in the locus of the matter distribution this argument is not conclusive. Moreover, as noted in \cite{static-typeI-bh} the causal structure of the ($t,r,x$) sector of the metric (\ref{ds-vacuum}) is analogous to the Bertotti-Robinson one, because they differ by an overall non-singular conformal factor, so radial null geodesics should behave similarly, also the equatorial ones. In fact the most significant point regarding the regularity of these spacetimes would be to clarify if any geodesic trajectory can hit some possible singularity, in particular in the presence of the black hole, not only the background. Actually this same issue might be worth scrutinizing for the seed itself: the black holes in the Bertotti-Robinson electromagnetic field. \\

\paragraph{About the spherical coordinates} We want to clarify a significative point regarding the two sets of spherical-like coordinates  $(t, r, x, \varphi)$ and $(t,\bar{r},\bar{x},\varphi)$. At this purpose we study the radial geodesics ($\dot{x}(\l)=0,\dot{\varphi}(\l)=0$) of the locally diffeomorphic metrics (\ref{BMMB-rx})-(\ref{BMMB-rx-fine}) and  (\ref{ds-vacuum}). Thanks to the conservation of the energy $E_0=-g_{tt}\dot{t}$, for static spacetimes, we can write the trajectory equation for a geodesics  with respect to the of affine parameter $\l$
\beq \label{geo}
                      g_{rr} \dot{r} + \frac{E_0^2}{g_{tt}} = \e
\eeq
where $\e$ is $1,0,-1$ for spacelike, lightlike or timelike geodesics respectively. Hence for the null equatorial radial geodesic (starting at the origin of the coordinates and directed towards infinity) relative to metric in coordinate ($\bar{r},\bar{x}$) we have
\beq \label{geo-rr}
                     \bar{ \D \l} \, = \, \frac{1}{E_0}\int_{0}^{\infty} \sqrt{-g_{tt} g_{\bar{r} \bar{r}}} \ d\bar{r} \ = \ \frac{1}{4 E_0 |B|} \left(2+ \frac{3\pi}{4} \right) \ < \ \infty \ .
\eeq
Hence  a photon\footnote{It is possible to show, regarding the metric (\ref{ds-vacuum}), that also for massive particles, that is $\e=-1$, an analogous result hold.}, travelling on a radial equatorial geodesics can reach $\bar{r}=\infty$ in a finite amount of affine parameter (for non zero values of $B$).\\ 
Similarly the coordinate time for a lightlike radial equatorial geodesic is characterised by a null line element
\beq
          0 = g_{tt} dt^2 + g_{\bar{r} \bar{r}}  d\bar{r}^2 \ ; 
\eeq
so the coordinate time for this radial equatorial geodesics is  
\beq
       \bar{ \D t } = \int_{0}^{\infty} \sqrt{-\frac{g_{\bar{r}\bar{r}}}{g_{\bar{t}\bar{t}}} }d\bar{r} = \frac{\pi}{2|B|}\ < \ \infty \ .
\eeq

On the other hand the same calculations about the affine parameter needed by a massless particle to reach the asymptotic boundary for the spacetime (\ref{BMMB-rx})-(\ref{BMMB-rx-fine}) along null, radial and equatorial geodesic gives 
\beq
        \D \l =  \infty \ .
\eeq

Therefore the ``radial coordinate'' $\bar{r}$ does not represent well the boundary of spacetime, while $r$ is better at this purpose. That's why the metric (\ref{ds-vacuum}) might be interpreted as geophysically incomplete, but it  seems more a coordinate issue than a physical problem. Thus $r$ seems better suited to describe the radial coordinate, at least, from a boundary perspective.\\

Similarly for $\e=1$ is easy to find that for large radial distances $r$ the affine parameter grows logarithmically 
\beq
         \D \l \ \propto \  r \ ,
\eeq
confirming that also in this case the $r$ coordinate can be considered a good radial coordinate.

For massive particles $ \e = -1 $, using the following normalisation for the tetra-velocity with respect to the affine parameter $\l$, the proper distance
\beq
       g_{rr} \left(\frac{dr}{d\l}\right)^2 + g_{tt} \left(\frac{dt}{d\l}\right)^2 = -1 \ ,
\eeq
and the above energy conservation we get, when $r>1/|B|$
\beq
          \left(  \frac{dr}{d \l}\right)^2 = - \frac{E^2+g_{tt}}{g_{tt} \ g_{rr}} \ = \ \frac{(4B^2r^2-1)(4E^2-B^2r^2)}{B^2 r^2} .
\eeq
Thus for $r>1/|B|$, since the above quantity must be positive, there is a maximum value for the radial coordinate $r_{max}$ that cannot be overcome by a massive particle, such that 
\beq
         r \le \frac{2 E}{|B|} = r_{max} \ .
\eeq
Therefore massive particle can travel at most to $r_{max}$, but then they must invert their trajectory. The proper time needed to reach $r_{max}$  from $r=1/|B|$ is 
\beq
       \D \l = \int_{1/|B|}^{4E^2} \frac{|B| r dr}{2 \sqrt{(B^2r^2-1)(4 E^2-B^2r^2)}} = \frac{\pi}{4 |B|} \ .
\eeq
Only massless physical particles can travel to infinity.\\
This analysis of the background in different coordinates holds similarly also in the presence of the black hole.\\

\paragraph{No conical singularities} The study of the conical singularities of the Reissner-Nordstrom-Bertotti-Robinson-Bonnor-Melvin metric (\ref{rn-brbm})-(\ref{AphiBRBM}) is particularly interesting, because the seed suffers from an angular defect when $\ell=0$. From a physical point of view this feature stems from the interaction between the intrinsic magnetic monopolar charge of the black hole and the external Bertotti-Robinson magnetic background. This interaction causes a ``force'' (from an intuitive Newtonian perspective) along the z axis, which tends to accelerate the black hole, even though the metric does not possess these extra acceleration features to describe this effect, such as an accelerating parameter. That's the reason the seed develops singular axial distribution of matter. A possible way to overcome these geometrical defects can be to find an equilibrium configuration between the relative position of the black hole and the external magnetic field, through the parameter $\ell$, as seen in section \ref{sec:RN-BR}. Alternatively one can generalise the metric introducing also the accelerating parameter, such as the one of the C-metrics, and fine tune it in order to get a regular configuration. This procedure has the advantage that leaves the original physical parameters unconstrained. A third  way, that  we pursue here, is to take into account also the interaction with the external Bonnor-Melvin magnetic field\footnote{A combination of all these possibilities can be pursuit simultaneously by considering accelerating charged metric in Bertotti-Robinson-Bonnor-Melvin external field and with a generic relative position between the black hole and external fields, so keeping non-zero all the parameters $A, B, b, \ell $. Note that the "force" is caused by the interaction of electromagnetic entities of the same kind: the magnetic monopole with the magnetic background. In case we work with electric monopoles in magnetic background, or vice versa an extra push along the z-axis cannot be realised (but this crossed interaction generates rotation).}. More specifically if  we compute the ratio between a small circle around the north and the south pole and the circle radius it is possible to detect conical singularities, whether  this ratio is not $2\pi$ on both sides simultaneously. In practice we have
\bea \label{cony-1}
           \lim_{x \to - 1} \frac{2\pi}{1-x^2} \sqrt{\frac{g_{\varphi\varphi}}{g_{xx}}}  & = & \frac{16 (1+B^2 m^2)^{3/2} (1+B^2 (m^2-e^2)) \D_\varphi}{\left(eB+\sqrt{1+B^2m^2}\right)^2 \left[4 + 4B^2m^2+be\left(be+4 \sqrt{1+B^2m^2}\right) \right]^2} \ , \\
           \lim_{x \to + 1} \frac{2\pi}{1-x^2} \sqrt{\frac{g_{\varphi\varphi}}{g_{xx}}}  & = & \frac{16 (1+B^2 m^2)^{3/2} (1+B^2 (m^2-e^2))\D_\varphi}{\left(eB - \sqrt{1+B^2m^2}\right)^2 \left[4 + 4B^2m^2+be\left(be - 4 \sqrt{1+B^2m^2}\right) \right]^2} \ .  \label{cony-2}    
\eea
The two quantities can be simultaneously $2\pi$ by constraining one parameter (for instance $b$) and setting $\D_\varphi$ properly
\bea
               b &=&  -2 \, \frac{1+B^2m^2\pm\sqrt{[1+B^2(m^2-e^2)](1+B^2m^2)}}{Be^2}  \ , \\
               \D_\varphi &=& \frac{4\sqrt{1+B^2m^2} \, [1+B^2(m^2-e^2)] \Big[\sqrt{1+B^2m^2} \mp \sqrt{1+B^2(m^2-e^2)} \Big]^2}{B^4e^4} \ . \nn
\eea
The role of the Bonnor-Melvin magnetic field, encoded into the integrating constant $b$ is fundamental not to have a trivial solution\footnote{The equality between (\ref{cony-1}) and (\ref{cony-2}) can be realised also for divergent values of $B$ or $b$. The resulting metric cannot be any more continuously connected to the Reissner-Nordstrom seed, but represents different regimes usually related to Levi-Civita asymptotic \cite{rotating-backgrounds}}. The two branches point to different regimes for finite values of the parameters. Of course the most relevant from a phenomenological perspective is related to finite and small intensities of the external magnetic fields (thus small $b$ and $B$), because the asymptotic deformation is not large and the spacetime represents a small deformation of spherical symmetric one.  Other particular regimes are possible, for instance the negative branch describes a spacetime void of conical singularities for $B=0$ when also $b=0$, so when we have not any background magnetic field but only the Reissner-Nordstrom black hole, while the positive branch indicate that the conical regularity is assured for large values of $B$ if $b\to\infty$.\\

\subsection{Special case: removal of the external magnetic field} 
\label{removal}

When the intrinsic electromagnetic charge $e=0$, it has been noticed, in \cite{bh+BRBM}, that a special fine tune\footnote{For the uncharged solution in the gauge of \cite{bh+BRBM} the constraint was $b=-B$. Different gauges may modify this value. In particular the gauge constant of the seed potential $A_{\varphi0}=-1/B$, as said above (\ref{Aphirx}), is useful for the vanishing $B$ limit. It gave in \cite{bh+BRBM} $b=-B$ (but actually there are always a couple of values for $b$, since the equation is quadratic in $b$. In this case the second value is not finite.). } between the intensities of the Bertotti-Robinson and the Bonnor-Melvin electromagnetic field (described by the parameters $B$ and $b$ respectively) allows one to remove the electromagnetic energy-momentum and obtain a static hole vacuum solution with an additional integration constant which deforms the background asymptotic (which is not flat), see \cite{static-typeI-bh} for more details about this spacetime. The Reissner-Nordstrom-Bertotti-Robinson-Bonnor-Melvin black hole built in (\ref{rn-brbm})-(\ref{AphiBRBM}) contains the natural generalisation of this solution. In this case, the main difference lays in the fact that the black hole is intrinsically charged by a monopole electromagnetic charge, so the fundamental properties of the black hole resemble more the Reissner-Nordstrom one rather the Schwarzschild one, especially close to the event horizon. For this reason the electromagnetic field cannot be completely removed for $e\ne 0$; clearly it is necessary to support the intrinsic electromagnetic charge of the black hole. Nevertheless a fine tuning between the Bertotti-Robinson parameter $B$ and the Bonnor-Melvin parameter $b$ can be imposed to elide the contribution of the external magnetic fields, similarly to the $e=0$ case. 
The condition that generalise the $e=0$ case can be derived by the requirement that the first order of the radial asymptotic expansion ($r \to \infty$) of the electromagnetic field is null: in practice we get
\beq \label{b-constraint}
                       b_\pm = \frac{2B \sqrt{1+B^2 m^2}}{A_{\varphi0} B \sqrt{1+B^2 m^2} \pm \sqrt{1+B^2(e^2+m^2) +2 B  \sqrt{1+B^2 m^2}}}
\eeq
In the vanishing monopolar charge subcase, i.e. $e \to 0$, the parametric constraint (\ref{b-constraint}) on the magnetic fields reduces to 
\beq \label{b-constraint-e0}
                      b_\pm  = \frac{2B}{B A_{\varphi0} \pm 1} \nn  \ .
\eeq
Of course that value is coherent with the result of \cite{bh+BRBM}. In fact, considering the constant gauge fixing (for the seed) $A_{\varphi0} = - 1/B $  we have $b = -B$ as in \cite{bh+BRBM}, \cite{static-typeI-bh}.\\

The constraint (\ref{b-constraint}) leaves the Dirac string, which is an indication of the monopolar magnetic charge of the metric, for $e \ne 0$. 
When the magnetic monopolar charge is null, that is for $e=0$, also the Dirac string disappears as expected (actually removing the whole electromagnetic field). There is also the possibility of removing the Dirac string imposing (\ref{b-constraint}) and at the same time by setting $A_{\varphi0}$, but this value makes the $b_\pm$ divergent, so these are peculiar scenarios addressed in \cite{rotating-backgrounds}.\\
Note that for small $B$ the constraint (\ref{b-constraint}) implies that also $b=0$, therefore the magnetic Reissner-Nordstrom (\ref{RN}) is retrieved. To have well defined limits it is better to confirm this fact by setting the gauge for the vector potential to $A_{\varphi0}=-1/B$.\\

When $e=0$ it is possible to analytically continue the solution by setting $B\to i B$, in this special case, to get a real vacuum metric describing a Schwarzschild black hole inside the expanding bubble of nothing \cite{bubble}, \cite{static-typeI-bh}, \cite{kerr+bubble}. But when $e \ne 0$ the metric and gauge potential seems not to have only real components, therefore the physical interpretation is unclear.

\section{Conclusions}
\label{sec:conclusion}

In this article we addressed the construction of the Reissner-Nordstrom black hole embedded in an external electromagnetic field composed by the combination of the Bertotti-Robinson and the Bonnor-Melvin electromagnetic field. To do so first we rotated the electric field of the Reissner-Nordstrom-Bertotti-Robinson to get a purely magnetic solution to use as a seed. The Petrov type of that solution has been identified as D. Then thanks to the Harrison transformation we added an extra Bonnor-Melvin electromagnetic field to get a magnetic Reissner-Nordstrom solution of Einstein-Maxwell equations, embedded into the Bertotti-Robinson-Bonnor-Melvin magnetic field. \\
Both the Bertotti-Robinson and the Bonnor-Melvin fields exert axial forces on the charged black hole, because of the interaction of the monopolar charge of the black hole and the external magnetic fields. The composition of the force allows the system to reach an equilibrium configuration for any position of the black hole, also in the simplest case, where it is in the centre of the coordinate system. This is not possible if only one of the two external fields is present. Therefore we have deduced the no conical singularity condition on the physical parameters, no string or strut are needed. Also the metric does not present curvature singularities in the main spherical coordinate patch examined. We noted that the same kind of metrics, although in a different coordinate system, might look geodetically incomplete because they reach $\bar{r}=\infty$ for a finite value of the affine parameter. They can be extended, but the analytical extension might encounter criticalities, which might be related to the coordinate chart more and to the appropriate definition of spatial infinity\footnote{Possible analogous issues could actually stems directly from the seed, Reissner-Nordstrom or even Schwarzschild embedded in the Bertotti-Robinson electromagnetic field, which, depending on the coordinate system considered, they also may need a similar analytical extension.}. The use of a proper set of spherical coordinates improves these issues. \\
We remark that during this generating process we were able to provide a diffeomorphism to relate different Petrov type D forms of the Schwarzschild back hole embedded in the Bertotti-Robinson external field, which has been a open problem lately.\\
Of course the natural generalisations could include the angular momentum to the model to get Kerr-Newman embedded in the Bertotti-Robinson-Bonnor-Melvin electromagnetic field in these spherical coordinates. It could be done in a straightforward way using, as a seed, the metric proposed in the convenient coordinate system of \cite{kerr-bertotti} or \cite{genuine-kerr-bertotti}. However since these solutions do not possess all the limits to their subcases clearly defined, for instance the limit to the Reissner-Nordstrom-Bertotti-Robinson remains unclear, it would be better to have the spacetime described by the coordinate system of this article.   \\
Further generalisations may also include the addition of the NUT parameter and the acceleration to have the complete family. The cosmological constant is more difficult to include because many solution generation techniques fail \cite{charging} if the action principle is modified.\\

\paragraph{Acknowledgements}
{\small I would like to thank Andrea Di Pinto and Adriano Vigan\`o for many stimulating discussions on this subject. A Mathematica notebook containing the solutions presented in this article can be found in the arXiv source folder.}\\

\vspace{0.1cm}

\appendix

\section{Electric Harrison transformation of the Schwarzschild-Bertotti-Robinson black hole}
\label{app:electric-harry}

In case we apply the Harrison transformation to the Schwarzschild-Bertotti-Robinson-Bonnor-Melvin black hole solution  \cite{bh+BRBM} we get surely a Reissner-Nordstrom spacetime embedded in an external electromagnetic field similar to the Bertotti-Robinson-Bonnor-Melvin one, however not exactly a combination of the two electromagnetic fields. That is because the Harrison transformation acting on the  Schwarzschild-Bertotti-Robinson-Bonnor-Melvin adds monopolar electric and/or magnetic charge to the black hole transforming the Schwarzschild geometry in the Reissner-Nordstrom one, but also the Harrison transformation acts on the  Bertotti-Robinson-Bonnor-Melvin background adding extra electromagnetic features. In fact the background can be considered as a couple of very far away charged black holes, as shown in the case  \cite{Emparan:2001gm}, therefore the Harrison transformation modifies the charges also of these sources of the electromagnetic field, so the electromagnetic field in between them.\\
To keep things as simple as possible, we show this mechanism by applying the Harrison transformation to add monopole charge to the Schwarzschild-Bertotti-Robinson. In this way we do not get the Reissner-Nordstrom black hole metric embedded into the Bertotti-Robinson electromagnetic field of type-$D$, such as the one in the sections  \ref{sec:RN-BR} or \ref{sec:RN-BRBM}. On the other hand  we get a different solution. Although it contains both the Reissner-Nordstrom black hole and the Bertotti-Robinson limits, it does not belong to the same Petrov class of the Reissner-Nordstrom-Bertotti-Robinson black hole of section \ref{sec:RN-BR}, in fact it belongs to the Petrov class I. The solution can be written 

\beq 
        ds^2 = g_{tt}(r,x) dt^2 + g_{rr}(r,x) \left[ \frac{dr^2}{\D_r(r)} + \frac{dx^2}{\D_x(x)} \right] - \frac{\D_r(r) \D_x(x) \,  \D_\varphi^2}{g_{tt}(r,x)} d\varphi^2 \ , 
\eeq
with
\bea
         g_{tt} &=& \frac{(r + B^2 r^3) [2 m  + B^2 m^2 r-r]}{\left[2 B p r (2 m - r + B^2 m^2 r) x + \left[r + p^2 (2 m - r + B^2 m^2 r)\right] \sqrt{1 + B^2 r \left[r + (2m + B^2 m^2 r-r) x^2\right]}\right]^2}   , \nn  \\
         g_{rr} &=& - \frac{\left(r^2 - 2 m r - B^2 m^2 r\right) \left(1 + B^2 r^2\right)}{g_{tt}\left[1 + B^4 m^2 r^2 x^2 + B^2 r \left(r + 2 m x^2 - r x^2\right)\right]^3} \ , \nn  \\
         \D_r &=&  \frac{(r^2-2mr-B^2m^2r) (1 + B^2 r^2)}{1 + B^2 r \left[r + ( 2m + B^2 m^2 r - r) x^2\right)} \ , \nn  \\
         \D_x &=& \frac{(1 - x^2) (1 + B^2 m^2 x^2)}{1 + B^2 r \left[r + ( 2m + B^2 m^2 r - r) x^2\right)} \ ,  \nn \\
         A_\varphi &=& \left\{2 m p x + \frac{1 + B^2 m r x^2 + p^2 \left[1 + B^2 m (2 r - m  + (2m + B^2 m^2 r -r) x^2)\right]}{B \sqrt{1 + B^2 r \left(r + (-r + m (2 + B^2 m r)) x^2\right)}}\right\} \Delta_\varphi \ . 
\eea
It has been used an imaginary parameter for the Harrison transformation to assure that the monopole charge added to the spacetime is of the same king of the one of the seed: magnetic in this case. In that way the metric remains diagonal. Choosing a real Lie-point parameter for the Harrison transformation generates a stationary rotating spacetime because of the interaction between the intrinsic electric charge of the black hole and the external magnetic field. The interested reader can also find this metric in the Mathematica notebook in the arXiv source files directory. Clearly the use of a general complex parameter generates a metric that is the superposition of the two: the black hole acquire  both intrinsic electric and magnetic monopole charges, in addition to the magnetic Bertotti-Robinson background of the seed.\\
To clearly identify the Reissner-Nordstrom black hole a shift in the radial coordinate is needed, for more details see \cite{Type-I}.
For instance, in the simple case where $B=0$, the radial shift and the reparametrization is given by
\beq
         r \to \frac{\bar{r}-2 p^2m}{1-p^2} \ , \hspace{1cm} t \to (1-p^2)\, \bar{t} \ , \hspace{1cm} m \to \frac{\bar{m}}{1+p^2} \ , \hspace{1cm} p \to \frac{\bar{m}-\sqrt{\bar{m}^2-\bar{p}^2}}{\bar{p}^2} \ .
\eeq

Similarly it is possible to build, with the same procedure, the electric Harrison transformation, a metric containing Reissner-Nordstrom and the Bertotti-Robinson-Bonnor-Melvin, from the Schwarzschild-Bertotti-Robinson-Bonnor-Melvin. However, for the same reasons above, when the Bonnor-Melvin parameter $b$ is null, the metric does not reduce to the type-$D$ Reissner-Nordstrom-Bertotti-Robinson. Hence it also possesses extra physical features. \\

\section{Schwarzschild-Bertotti-Robinson in spherical coordinates}
\label{app:SBR-rx}

For seek of completeness we write explicitly the Schwarzschild-Bertotti-Robinson solution in the spherical coordinate $(r,x)$, that is the $e=0$ limit of the Reissner-Nordstrom-Bertotti-Robinson of section \ref{sec:RN-BR-spherical}. This solution was first found in \cite{alekseev-garcia}, in a different set of coordinates. The solution can be written as 
\beq \label{s-br-rx-inizio}
        ds^2 = g_{tt}(r,x) dt^2 + g_{rr}(r,x) \left[ \frac{dr^2}{\D_r(r)} + \frac{dx^2}{\D_x(x)} \right] - \frac{\D_r(r) \D_x(x) \D^2_\varphi}{g_{tt}(r,x)} d\varphi^2 \ , \nn
\eeq
with
\bea
          g_{tt} &=& -\frac{\left[1 + B^2 (-2 m r + r^2 + m^2 x^2) + H(r, x)\right] \left[2 m - 2 r + m \sqrt{2} \sqrt{J(r, x)}\right]^2}{8 r (-2 m + r)} \nn \\
          g_{rr} &=& \frac{\left(1 + B^2 m^2\right) \left[-2 m + 2 r + m \sqrt{2} \sqrt{J(r, x)}\right]^2}{4 H(r, x)} \nn  \\
          H(r,x) &=& \sqrt{4 B^2 \left(1 + B^2 m^2\right) r (-2 m + r) x^2 + \left(1 + B^2 \left(2 m r - r^2 + m^2 x^2\right)\right)^2} \ , \nn \\
          J(r,x) &=& 1 - B^2 \left(-2 m r + r^2 + m^2 x^2\right) + H(r, x) \ , \nn \\
          \Delta_r(r) &=& r^2-2mr \ , \hspace{2cm} \Delta_x(x) = 1-x^2  \ , \nn \\ 
          A_\varphi &=&\frac{4 B^2 m (m - r) + \frac{8 B^2 m (m - r) x^2}{-J(r, x)} + \frac{4 \sqrt{2} B^2 (m - r)^2 x^2}{[J(r, x)]^{3/2}} + \frac{2 \sqrt{2} \left[1 + B^2 (2 m r - r^2 + m^2 x^2)\right]}{\sqrt{J(r, x)}} + \sqrt{2} \sqrt{J(r, x)}}{4 B \left(1 + B^2 m^2\right)} \D_\varphi \nn \ .
\eea
As usual the function $g_{rr}$ is defined up to a constant dilatation and $\D_\varphi$ is introduced to adjust possible angular deficit on the azimuthal axis.\\
It could be useful to provide the change of coordinates from the above metric to obtain the Schwarzschild-Bertotti-Robinson in the coordinates $(\bar{r},\bar{\theta})$, as in (\ref{schwarschild-br-spherical}).
\bea
      r &=&  m + \frac{\bar{r} - m}{\sqrt{1 + B^2 \bar{r} \left\{\bar{r} + \left[-\bar{r} + m \left(2 + B^2 \bar{r} m\right)\right] \cos^2 \bar{\theta} \right\}}} \ , \nn \\
      x &=& \frac{\left(1 + B^2 m \bar{r}\right) \cos \bar{\theta} }{\sqrt{1 + B^2 \bar{r} \left\{\bar{r} + \left[-\bar{r} + m \left(2 + B^2 \bar{r} m\right)\right] \cos^2 \bar{\theta} \right\}}}
\eea
The same map can be used to obtain the alternative electric potential $A_t$ from (\ref{Atrx}) setting $e=0$.

\end{document}